\documentclass[a4paper,11pt]{article}
\begin{document}
\title{IRREVERSIBILITY IN CLASSICAL MECHANICS}
\author{V.M. Somsikov}
\date{\it{Laboratory of Physics of the geoheliocosmic relation,\\
Institute of Ionosphere, Almaty, 480020, Kazakhstan\\E-mail:
nes@kaznet.kz}} \maketitle
\begin{abstract}
An explanation of the mechanism of irreversible dynamics was
offered. The explanation was obtained within the framework of laws
of classical mechanics by the expansion of Hamilton formalism.
Such expansion consisted in adaptation of it to describe of the
non-potential interaction of a systems. The procedure of splitting
of a system into equilibrium subsystems, presentation of
subsystem's energy as the sum of energy of their relative motion
and their internal energy was the basis of the approach which was
used for the analysis of nonequilibrium systems. As a results the
generalized Liouville equation and equation of subsystems
interaction was obtained. Based on these equations, the
irreversible transformation of energy of the relative motion of
subsystems into their internal energy was proved. The formula
which expresses the entropy via the work of subsystems'
interaction forces was submitted. The link between classical
mechanics and thermodynamics was discussed.
\end{abstract}

\section{Introduction}

Collective properties of natural systems should be connected with
laws of dynamics and properties of their elements. Searching of
these connections is among the main tasks of physics. There are
principal difficulties in connection with solving of this problem.
Particularly they are consisted in absence of links between laws
of classical mechanics and laws of the thermodynamics. The
irreversibility problem which was formulated by Boltzmann about
150 years ago is the key problem among them [1-4]. The essence of
this problem is that dynamics of natural systems is irreversible,
while in accordance with Newton equation dynamics of elements of
system is reversible.

Actually all attempts of irreversibility substantiation within the
framework of canonical formalism of Hamilton were somehow or other
stricken against Poincare's theorem about reversibility. This
theorem forbids the existence of irreversibility within
Hamiltonian systems [5-8].

Study of irreversibility is often based on the Lorentz's gas model
[7-9]. This model represents point objects with certain masses.
They are elastically scattered on various boundaries but do not
interact between each other. Study of Lorentz's gas allowed to
find important laws of behavior of dynamic systems, and mixing is
among them. But nevertheless the problem of irreversibility has
remained unsolved. It was stricken against so-called "coarse
grain" of the phase space problem [8].

At the beginning of our irreversibility investigation we chose the
most elementary system - system of hard disks. Taking collisions
of disks among themselves into account was our first step in study
of irreversibility. The motion equation of colliding hard disks
was obtained. Based on it we found out that forces of interaction
of disks systems depend on its relative velocities, i.e. they are
non-potential [10-11]. But in order to study such systems we had
to expand the framework of Hamilton canonical formalism in such a
way that it would be possible to use it for description of
dynamics of systems with non-potential forces (polygenic forces
[12]). Therefore Lagrange, Hamilton and Liouville equations for
systems with non-potential forces acting between them were
obtained [11, 15]. According to accepted terminology [12], we will
call these equations and system-system interaction forces as {\it
"generalized equations"} and {\it "generalized forces"}.

The analysis of generalized Liouville equation led us to
conclusion that irreversibility is possible only for
non-potentially interacting systems. Using an equation of disks'
motion and generalized Liouville equation we were succeeded in
showing of existence of irreversible dynamics for the
nonequilibrium system of disks which is presented in the form of
two colliding subsystems. It was found out that the
irreversibility is determined by decreasing of generalized forces
[11].

The generality of results of researches of disks systems is
limited to a condition of their absolute rigidity. Moreover, a
series of assumptions at the proof of irreversibility have been
used. It has essentially simplified the proof of irreversibility
for systems of the disks. Therefore for generalization of these
results on a real systems it was necessary to find such approach
which would allow to remove the restriction connected with these
assumptions and assumption about non-potentiality forces between
elements of the system.

For this purpose we took into account that the presence of
dependence of the systems interaction forces from velocities is
the necessary condition for existence of irreversibility. This
result allowed to reduced the problem concerning irreversibility
of natural systems to the problem whether generalized forces
acting between interacting systems are non-potential when forces
acting between their elements are potential.

For analyzing generalized forces the fact that usually
nonequilibrium systems can be represented in the form of set of
equilibrium subsystems which are in motion relative to each other
was used [2, \S10]. It gave possibility to reduce problem of
irreversible dynamics to the proof of non-potentiality of
generalized forces and that these forces will be decreased. This
proof was carried out in such way. In homogeneous space a closed
nonequilibrium system of potentially interacting elements was
prepared. This system was divided into a set of equilibrium
subsystems. The energy of subsystem was presented in the form of
the sum of internal energy, energy of relative motion and an
interaction energy of the subsystem. After that the equation for
energies' exchange between subsystems was obtained. In agreement
with this equation we found out that kinetic energy of subsystems
will be transformed both to potential and internal energy of
subsystems. Then we showed that reverse transformation of internal
energy of subsystems to energy of their motion is forbidden by the
law of conservation of momentum. It provides relaxation of the
system to equilibrium.

The connection of classical mechanics with thermodynamics follows
from the fact that the work of generalized forces will change not
only the energy of motion of a subsystem as a whole, but also its
internal energy.

The fact that energy of subsystems' motion is irreversibly
transformed to their internal energies allowed us to obtain the
formula for the entropy deviation of the non-equilibrium system
from equilibrium value.

Here a results of our investigation of the mechanism of
irreversibility and interconnection between classical mechanics
and thermodynamics are offered.

\section{The generalized Liouville equation and irreversibility of a hard-disks system}

The equation of motion for hard disks is deduced on the basis of
collision matrix from a laws of conservation of energy and a
momentum. This equation can be written as [17]:
\begin{equation}
\dot{V}_k=\Phi_{kj}\delta (\psi_{kj}(t))\Delta_{kj}\label{eqn1}
\end{equation}
where ${V_{k} = V_{x}+iV_{y}}$ is a complexity form of velocity of
$k$-disk along $X$ and $Y$ arcs consequently;
$\psi_{kj}=[1-|l_{kj}|]/|\Delta_{kj}|$;
$\Phi_{kj}=i(l_{kj}\Delta_{kj})/(|l_{kj}||\Delta_{kj}|)$;
$\delta(\psi_{kj})$ is a delta function; $l_{kj}(t)=z_{kj}^0+\int
\limits_{0}^{t}\Delta_{kj}{dt}$ are distances between centers of
colliding disks; $z_{kj}^0=z_k^0-z_j^0$, $z_k^0$ and $ z_j^0$ -
are initial values of disks coordinates; $k$ and $j$ are numbers
of colliding disks; $i$ is an imaginary unit; $t$ is a time;
$\Delta_{kj}=V_k-V_j$ are relative disks' velocities; $D$ is
disks' diameter. Collisions are considered to be central, and
friction is neglected. Masses and diameters of disks are accepted
to be equal to 1. The moments of $k$ and $j$ disks collisions are
determined by $\psi_{kj}=0$ equality.

The force of disks' interaction, which is determined by the right
side of eq. (1), depends on relative velocities of colliding disks
and impact parameters. Therefore it is not a gradient of scalar
function. As a force acting between any pair of disks depends on
velocity, i.e. it is non-potential then forces acting between
subsystems of disks are non-potential too and depend on
velocities. Therefore for description of there dynamics it is
impossible to use canonical equations of Lagrange, Hamilton and
Liouville [12-14]. Hence, it is necessary to replace these
equations by generalized ones which at least are applicable to the
open systems with non-potential forces.

The generalized Liouville equation for subsystems was obtained in
following way [15, 17]. We took a closed nonequilibrium system
which consists of $N$ elements. We divided this system into $R$
equilibrium subsystems. Then we selected one of subsystems, which
was called as $m$-subsystem. With the help of D'Alambert equation,
the generalized Lagrange and Hamilton equations were obtained for
$m$-subsystem at the basis of variational method. At obtains of
these equations we considered, that forces between subsystems
generally non potential. On the basis of these equations the
generalized Liouville equation was obtained. This equation can be
written as [11,15]:

\begin{equation}
{\frac{df_m}{dt}=-f_m
\frac{\partial}{\partial{\vec{p}_k}}\vec{F}^m} \label{eqn2}
\end{equation}

Here $f_m=f_m({\vec{r}}_k,{\vec{p}}_k,t)$ is a normalized
distribution function for $m$-subsystem elements; ${\vec{F}^m}$ is
generalized force acting on $m$-subsystem,
$\vec{F}^m=\sum\limits_{k=1}^L\sum\limits_{s=1}^{N-L}\vec{F}_{ks}^m$;
${k=1,2...L}$ are disks of $m$-subsystem; $s= 1, 2, 3,..., N-L$
are external disk acting on $k$-disk of $m$-subsystem; $m=1, 2,
3,...R$; ${\vec{p}}_k$ and ${\vec{r}}_k$ are momenta and
coordinates of $m$-subsystem disks consequently.

The eq.(2) describes change of particles distribution function of
the selected equilibrium subsystem on condition that non-potential
forces are acting on it. This equation essentially differs from
canonical Liouville equation. Actually canonical Liouville
equation is suitably for description of potentially interacted
systems on conditions for short enough intervals of time when it
is possible to neglect an exchange of energy between systems [2].

The important step during generalization of Liouville equation was
to refuse requirement of potentiality of forces acting between
subsystems. It makes this equation applicable for study of
irreversibility [11].

Let us note that the similar form of the generalized Liouville
equation can be obtained if the forces is accepted to be
dissipative [16]. But in this case generality of obtained equation
would be lost. It is so because postulating of dissipative forces
is equivalent to postulating of irreversibility. Moreover
acceptance dissipative forces essentially narrows field of
applicability of this equation. Indeed, forces of interaction of
disks' systems are nondissipative though they are non-potential
[17].

A physical essence of the right hand side of the eq. (2) that it
is the integral of collisions which can be obtained with the help
of the motion equations of elements. Both under condition of
potentiality of forces, and equilibrium condition when relative
motion of subsystems is absent, the right side of the eq. (2) is
equal to zero.

Let us show, how it is possible to obtain possibility of existence
of the irreversible dynamics under condition of non-potentiality
forces acting between subsystems using the generalized Liouville
equation [10].

Because the equality $\sum\limits_{m=1}^{R}\vec{F}^m =0$ is
correct then the following equation for Lagrangian of entire
system, $L_R$,
${\frac{d}{dt}\frac{\partial{L_R}}{\partial{\vec{v}_k}}-
\frac{\partial{L_R}}{\partial{\vec{r}_k}}=0}$ and the appropriate
Liouville equation:
${\frac{\partial{f_R}}{\partial{t}}+{\vec{v}_k}\frac{\partial{f_R}}
{\partial{\vec{r}_k}}+\dot{\vec{p}_k}\frac{\partial{f_R}}{\partial{\vec{p}_k}}=0}$
will have a place. Here $f_R$ is distribution function which
corresponds to entire system; $\vec{v}_k$ is a velocity of
$k$-disk. The entire system is conservative. Therefore we will
have: ${\sum\limits_{m=1}^R div\vec{J}_m=0}$. Here
$\vec{J}_m=(\dot{\vec{r}_k},{\dot{\vec{p}_k}})$ is a generalized
current vector of the $m$-subsystem in a phase space. This
expression is equivalent to the following equality:
${\frac{d}{dt}(\sum\limits_{m=1}^{R}\ln{f_m})}=
\frac{d}{dt}(\ln{\prod\limits_{m=1}^{R}f_m})=
{(\prod\limits_{m=1}^{R}f_m)}^{-1}\frac{d}{dt}(\prod\limits_{m=1}^{R}{f_m})=0$.
So, $\prod\limits_{m=1}^R{f_m}=const$. For equilibrium state we
would have $\prod\limits_{m=1}^R{f_m}=f_R$. Because the equality
$\sum\limits_{m=1}^{R}\vec{F}^m=0$ is fulfilled at all times, we
have the equality, $\prod\limits_{m=1}^R{f_m}=f_R$, as the
integral of motion. This is in agreement with Liouville theorem
about conservation of a systems' phase space [2].

So, Liouville equation for entire non-equilibrium system is in
accordance with the generalized Liouville equation for selected
subsystems only in two cases: if the condition that
$\int\limits_{0}^{t}{(\frac{\partial}
{\partial{\vec{p}_k}}\vec{F}^m)}dt\rightarrow{const}$ is satisfied
when $t\rightarrow\infty$, or when
$\frac{\partial}{\partial{\vec{p}_k}}\vec{F}^m$ is a periodic
function of time. The first case corresponds to irreversible
dynamics, and the second case corresponds to reversible dynamics.

Thus irreversible dynamics is possible under condition of
redistribution of phase-space volume between subsystems while full
volume is an invariant. Reversibility exists if the system is
placed near to equilibrium or it is located in cyclic points of
phase-space. In the second case a periodic change of phase-space
volume of subsystems takes place under condition of system
phase-space volume preservation as a whole [9].

Thus generalized Liouville equation allows to describe dynamics of
nonequilibrium systems within the framework of classical
mechanics. According to this equation both reversible and
irreversible dynamics take place. Irreversibility is possible only
if motions of subsystems are presented and generalized forces
depend on subsystems' velocities. Presence of such dependence
eliminates an interdiction of irreversibility which is dictated by
the Poincare's theorem about reversibility. Therefore to prove the
existence of irreversibility first of all it is necessary to show
presence of the relative motion of subsystems in nonequilibrium
systems and non-potentiality of forces acting between subsystems.

For hard disks the irreversibility is possible because forces
acting between subsystems are non-potential (necessary condition).
Therefore if we show that these forces aspire to zero
(sufficiently condition) then irreversibility for them will be
proved.

Let's take the nonequilibrium system of disks which consists of
from two equilibrium subsystems: $L$ and $K$ consequently. Let
$L$- subsystem fly at $K$- subsystem. Let's designate velocity of
the center of mass $L$-subsystem as $\vec{V}_L$. Let's assume that
all disks collide simultaneously in and short enough intervals of
time $\tau$. Such assumption does not have influence on
qualitative characteristics of evolution. Then the equation of
motion of disks (4) will be as follows [11]:
\begin{equation}
\dot{\vec{V}}_k^n=\Phi_{kj}^n\Delta_{kj}^{n-1}\label{eqn3}
\end{equation}

Here $k=1, 2, 3,...L$ is $L$-subsystem disks' numbers. $j$ number
corresponds to each $k$ and point of time $t=n{\tau}$ there
corresponds number $j$; ${{k}{\neq{j}}}$.

Taking accepted requirements into account it is possible to obtain
the expression for the force acting on a subsystem by summing the
equations (3) for all disks of $L$-subsystem. After such summation
we will obtain:

\begin{equation}
\dot{\vec{V}}_L^n=
\sum\limits_{k=1}^L\Phi_{ks}^n\Delta_{ks}^{n-1}\label{eqn4}
\end{equation}

Here we have taken into account that collisions of disks
$L$-subsystem with external $s$ - disks make contribution to the
right side of eq. (4). This equation is written down in approach
of pair collisions.

The eq. (4) describes change of a total momentum, effecting onto
the  $L$-subsystem as a result of collisions at the moment
$n\tau$. The aspiration of a total momentum to zero is equivalent
to aspiration to zero of the force, acting on the part of external
disks. Thus the eq. (4) determines changes of relative velocities
of subsystem .

When $L\rightarrow\infty$ due to mixing the uniformity of
distribution of impact parameters of disks takes place. Really, in
accordance with the mixing condition we have the following [13]:
$\mu(\delta)/\mu(d)=\delta/d$ (a). Here $\mu(\delta)$ is a measure
which corresponds to the total value of impact parameter "$d$";
$\delta$ is an arbitrary interval of impact parameters and
$\mu(\delta)$ is corresponding measure. The fulfillment of (a)
condition  means the proportionality between the number of
collisions of disks which fall at the $\delta$ interval, and disk
diameter-$d$. I.e. the distribution of impact parameters is
homogeneous.

If (a) is correct then condition of decay of correlations is
right. Therefore in accordance with eq. (4):
$<\Phi_{ks}^n\Delta_{ks}^{n-1}>=<\Phi_{ks}^n><\Delta_{ks}^{n-1}>$.
As the first multiplier depends on impact parameters, and the
second one depends on relative velocities the condition of decay
of correlations is equivalent to condition of independence of
coordinates and momentum [3, 13]. I.e., when $L\rightarrow\infty$
it is possible to turn from summation to integration of phase
multiplier $\phi=<\Phi_{ks}^n>$ on impact parameter. Then we will
have [11]: $\phi=\frac{1}{L}\lim\limits_{L\rightarrow\infty}\
\sum\limits_{k=1}^L\Phi_{ks}^n=
\frac{1}{G}\int\limits_0^{\pi}\Phi_{ks}^nd(\cos\vartheta)=-\frac{2}{3},$
where $G=2$ is normalization factor; $(\cos\vartheta)=d$ is impact
parameter.

As relative velocities of subsystems are determined by velocities
of centers of the mass we will have:
$\vec{V}_L^n=<\Delta_{ks}^{n-1}>$. Therefore, we have:
$\vec{V}_L^n=-\frac{2}{3}\vec{V}_L^{n-1}$ (b).

Thus velocity of a subsystem is decreased. The rate of decreasing
is determined by 2/3 factor. The reason of such decreasing is
related to chaotization of vectors of disks' velocities. Thus, we
have obtained that the nonequilibrium disks' system is
equilibrate.

Let us note that mathematical operations used by us are not
connected neither with averaging of phase space nor with any other
change of real nature of phenomenon which is under investigation.

Averaging of the phase space is usually used during
irreversibility investigation on the basis of analysis of
behaviour of phase trajectory. In the mixing systems phase
trajectory eventually fills phase space very densely, it is
infinitely closed to any accessible point. The irreversibility
relates to infinite fine splitting of the phase space by phase
trajectory. It can be performed  by averaging of the phase space
i.e. by "coarse grain". But as a result of such operation the
motion of phase trajectory becomes probabilistic and therefore
irreversible [8]. I.e. operation of averaging of the phase space
introduces irreversibility into a system. As the nature of the
"coarse grain" of phase space is unknown and, most likely, there
is no such nature, this approach does not solve a problem of
irreversibility.

We are searching for the mechanism of irreversibility by analyzing
of the generalized force. Though this dynamic parameter is
collective and equals sum of exterior forces acting on elements of
subsystem, but it is strictly determined equation of motion also
as well as a phase trajectory. During the analysis of the force we
also had used mixing property. It has allowed us to use
integration of forces instead of summation in thermodynamic limit.
But in our case integration was carried out only for estimation of
value of change of generalized force due to disks' collisions.
Thus physical nature of the process was not distorted.

Let us note that according to offered mechanism of irreversibility
the equilibrium state is asymptotically stable. I.e. on deviation
of system from an equilibrium point the system returns back to it
[11]. Really, on deviation of system from equilibrium the
generalized force $\vec{F}_L$ will appear. As it was shown (see
eq. (b) ) this force is a decreasing force. Therefore equilibrium
is asymptotically stable.

It is not difficult to determine characteristic time which is
needed for system to return back to equilibrium on deviation from
it. In accordance with expression (b) it can be written down as
$t_{din}\sim\int\frac{d\vec{V}_L}{\vec{F}_L}$. I.e., if the system
leaves equilibrium due to exterior action then it will return back
in characteristic time $t_{din}\sim\frac{1}{|F_L|}$.

Important conclusion follows from here. It consists in the fact
that stability of equilibrium point causes restriction of probable
fluctuations. Really, let's suppose that the system has deviated
from equilibrium in a probable way. A time of deviation is
determined by the probability laws. According to Smoluchowski
formula [8, 22], for ergodic systems average resetting time,
$t_p$, or Poincare's cycle time is equal to
$\tau=t_p(1-P_0)/(P_0-P_1)$, were $P_1$ -is a probability of
return in $t_p$ time; $P_0$ is a probability of initial phase
region. But on deviation of system from equilibrium returning
force $\vec{F}_L$ appears. This force becomes bigger as system
becomes farther from equilibrium. Therefore the amplitudes of
possible fluctuations is restricted by this force. I.e.
spontaneous deviation of system from equilibrium state is possible
only for such points of phase space for which characteristic time
$t_{din}$ which is determined by field of generalized forces is
more than characteristic time $\tau$ which is necessary for
spontaneous deviation. This condition determines the framework of
applicability of probabilistic description of dynamics of systems
with the mixing [10, 11].

Thus, in agreement with the generalized Liouville equation the
dependence system-system interaction forces from their relative
velocities is a necessary requirement of occurrence of
irreversible dynamics. As pair interactions of disks depend on
their velocities, the disks system-system interaction forces also
depend on velocities. It causes an opportunity of existence of
irreversibility. Such opportunity is realized because of decrease
of forces of interaction of equilibrium subsystems of disks into
which the nonequilibrium system is divide. In a result its system
is equilibrates.

The conclusion about irreversibility for a hard-disks system will
confirm further at the analysis of the more general case of
dynamics of nonequilibrium systems of potentially interacting
elements. For this purpose, using the procedure of splitting a
system into equilibrium subsystems, presentation of their energies
as the sum of their energy of motion and internal energy, the
equation of interaction of such subsystems will be obtained. Based
on it we will show, that the reason of irreversible dynamics in
nonequilibrium system of potentially interacting elements are a
non-potentiality  of the generalized forces of interaction of
subsystems and decreasing of them.

\section{The relative motion of subsystems}

From generalized Liouville equation follows, that irreversibility
is possible only when the forces of interaction of equilibrium
subsystems are non-potential. Therefore it is clear, that if
nonequilibrium system represents a set of equilibrium subsystems,
which is in relative motion, the proof of irreversibility is
reduced to the proof of decreasing of their relative velocities
that is equivalent to decreasing of system-system interaction
forces.

Representation of a nonequilibrium system in the form of set of
selected equilibrium subsystems corresponds to local equilibrium
approach which is used in statistical physics. Possibility of such
representation is explained  by the fact that the time of
equilibration for a subsystem is much less than that for all
system [2, 3]. Let's show that such equilibrium subsystems will be
characterized by relative motion. This fact will allow to reduce
the proof of irreversibility for any systems to the proof of
decrease of generalized forces without decrease of generality.

The proof of existence of relative motion of subsystems for
nonequilibrium systems follows from [2, $\S 10$]. Let's take the
nonequilibrium system of disks which consists of $R$ equilibrium
subsystems. In accordance with [2] the entropy $S$ for the system
can be written down as:
\begin{equation}
S={\sum\limits_{m=1}^R}S_m(E_m-T_m^{tr})\label{eqn5}
\end{equation}
Here $S_m$ is entropy for $m$-subsystem; $E_m$ is full energy of
$m$-subsystem; $T_m^{tr}=P_m^2/2M_m$ is a kinetic energy of motion
of $m$-subsystem; $M_m$ is subsystem's mass; $P_m$ is momentum.
The argument of $S_m$ is internal energy of $m$-subsystem.

As the system is closed we have:
${\sum\limits_{m=1}^R}\vec{P}_m=const$,
${\sum\limits_{m=1}^R}{[\vec{r}_m\vec{P}_m]}=const$. Here
$\vec{r}_m$ is a position vector of $m$-subsystem.

In equilibrium state the entropy as a function of momentum of
subsystems is maximum. It is possible to determine necessary
conditions of maximum using method of uncertain Lagrange
multipliers; momentum derivatives from the following expression
should be equal to zero:
\begin{equation}
{\sum\limits_{m=1}^R}\{S_m+aP_m+b[r_mP_m]\},\label{eqn6}
\end{equation}
where $a, b$ are constant multipliers.

If we differentiate $S_m$ with respect to $P_m$ taking definition
of temperature into account we will obtain: $\frac{\partial}
{\partial{P_m}}S_m(E_m-T_M^{tr})=-P_m/(M_mT)=-v_m/T$. Hence,
differentiating eq.(5) with respect $P_m$, we shall have:
$v_m=u+[{\Omega}r_m]$ (a), where ${\Omega=bT}, u=aT$, $T$ is a
temperature. From this fact follows that the entropy is maximum
when velocities of all subsystems are determined by the formula
(a). According to this formula all subsystems should move with
identical translational velocities and to rotate with identical
angular velocity in equilibrium. It means, that a closed system in
equilibrium state can only move and rotate as the whole; any
relative motions of subsystems are impossible.

As it follows from the formula (5), the rate of system's deviation
from equilibrium is determined by $T_m^{tr}$ value. This energy
can be selected by dividing the system into equilibrium
subsystems. In accordance with eq. (5) the process of
equilibration is caused by transformation of energy $T_m^{tr}$ to
internal energy of systems.

Thus, equilibrium is characterized by condition that $T_m^{tr}
=0$, which have a place on any splitting of the equilibrium system
into subsystems. Therefore subsystems will have relative motion in
nonequilibrium system. If the system goes to equilibrium then
$T_m^{tr}$ energy should aspire to zero.

\section{The subsystems dynamics}

The proof of irreversibility for hard disks is based on dependence
of velocity on disks' interaction forces. But all fundamental
forces in the nature are potential [20]. And according with
generalized Liouville equation the presence of dependence of the
generalized forces on velocities of motion of subsystems is
necessary for existence of irreversibility in systems of
potentially interacting elements. Therefore the problem concerning
irreversibility of natural systems is reduced to the problem
whether forces acting between systems are non-potential when
forces acting between their elements are potential. It will be
shown below that such dependence have place for nonequilibrium
systems of potentially interacting elements.

We will solve this problem in such way. Let's take a system of
potentially interacting elements which was prepared as
nonequilibrium. We represent it in the form of set of equilibrium
subsystems. After that we obtain the equation for energies'
exchange between subsystems. We will search for this equation
using requirement of existence for subsystems of not one, but two
forms of energies  - energies of their motion as a whole and their
internal energy. The energy of interaction of systems will be
introduced into the right part of the equation. The energy of
subsystem we present in the form of the sum of its energy of a
motion, an internal energy and an interaction energy. Internal
energy will be represented in the form of the sum of a kinetic
energy of motion of system's particles concerning its centre of
mass and a potential energy of interaction of particles inside of
subsystem. Then time derivation from the energy presented thus
will give the required equation which describes transformation of
energy between subsystems. If we will obtain non-potentiality of
the forces acting between subsystems from this equation then
according to generalized Liouville equation the irreversible
dynamics is possible.

Thus, as well as in the [18], we use a model of a simple system of
particles. The basic distinction  that our approach is
non-statistical. It allows to determine the nature of
nonconservative forces within the framework of classical mechanics
and to analyze connection of the classical mechanics with the
first and the second laws of thermodynamics.

Let's take the system which consists of $N$ elements. Masses of
elements are accepted to be equal to $1$. We will present energy
of the system as the sum of kinetic energy of motion of the system
as a whole-$T_N^{tr}$ and the kinetic energy of the motion of its
elements concerning the center of mass- $\widetilde{T}_N^{ins}$;
and their potential energy- $\widetilde{U}_N^{ins}$. The energy
$E_N^{ins}=\widetilde{T}_N^{ins}+\widetilde{U}_N^{ins}$ is
internal energy of the system. It equals the sum of the kinetic
energy of relative motion of elements and the energy of potential
interaction. The $T_N^{tr}$ energy is determined by $\vec{V}_N$
velocity of the center of mass. Therefore this energy is
characterise the level of regularity of particles' velocities
because $\vec{V}_N=\frac{1}{N}\sum\limits_{i=1}^N\vec{v}_i$ .

When external forces are absent, $T_N^{tr}$ and $E_N^{ins}$
energies are constants. Due to the momentum preservation law these
energies are the motion integrals.

The full energy of closed system of potentially interacting
elements in homogeneous space can be presented as:
$E_N=T_N+U_N=const$, where
$T_N=\frac{1}{2}\sum\limits_{i=1}^N{{\vec{v}_i}^2}$ is a kinetic
energy; $U_N(\vec{r}_{ij})$ is a potential energy;
$\vec{r}_{ij}=\vec{r}_i-\vec{r}_j$ is the distance between $i$ and
$j$ elements.

The equation of motion for elements of system can be obtained by
means of differentiation of expression of energy of systems with
respect to time [12]. We will have:
$m\dot{\vec{v}}_i=-\sum\limits_{i=1,j\neq{i}}^N\vec{F}_{ij}$,
where $\vec{F}_{ij}=\frac{\partial} {\partial{\vec{r}_{ij}}}U$
(c). It is a Newton equation .

Owing to splitting of nonequilibrium system into equilibrium
subsystems, the problem concerning irreversibility nature is
reduced to the problem concerning character of subsystems' energy
exchange. We will call the internal energy of equilibrium
subsystem as {\it "bound energy"} to emphasize the absence of
energy of relative motion of microsystems on which this subsystem
can be divided.

To simplification let's take the system which consists of two
interacting equilibrium subsystems. They are $L$ and
$K$-subsystems. $\vec{V}_L$ and $\vec{V}_K$ are velocities of the
centers of mass of corresponding subsystems. Number of elements in
$L$-subsystem is equal to $L$, and number of elements in $K$
-subsystem is equal to $K$. Let us suppose that $L+K=N$ and
$L\vec{V}_L+K\vec{V}_K=0$, i.e. the center of mass of system is
motionless.

Equations for energy exchange between subsystems can be obtained
by means of differentiation of system energy with respect of time
along with grouping terms which correspond to elements of
different subsystems together.

Using summation of changes of kinetic and potential energies of
elements in each subsystem we will obtain for $L$-subsystem:
${\sum\limits_{{i_L}=1}^L}m\vec{v}_{i_L}{\dot{\vec{v}}_{i_L}}$+
${\sum\limits_{j_L=i_L+1}^L}{\sum\limits_{i_L=1}^{L-1}\vec{v}_{{{i_L}{j_L}}}}{\vec{F}_{{i_L}{j_L}}}$
= $-{\sum\limits_{j_K=1}^K}\sum\limits_{i_L=1}^{L}\vec{v}_{{i_L}}$
$\vec{F}_{{i_L}{j_K}}$ (d), and for $K$-subsystem:
${\sum\limits_{{i_K}=1}^K}m\vec{v}_{i_K}{\dot{\vec{v}}_{i_K}}+
{\sum\limits_{j_K=i_K+1}^K}{\sum\limits_{i_K=1}^{K-1}\vec{v}_{{{i_K}{j_K}}}}$
$\vec{F}_{{i_K}{j_K}}$=
${\sum\limits_{j_K=1}^K}\sum\limits_{i_L=1}^{L}\vec{v}_{{j_K}}$
$\vec{F}_{{i_L}{j_K}}$ (e). The subindexes $K$ and $L$ denote the
subsystem to which elements belong.

Let's transform the left side of the eqs. (d, e) using following
equality:
${\sum\limits_{{i}=1}^N}m\vec{v}_{i}{\dot{\vec{v}}_{i}}$=$N\vec{V}_N{\dot{\vec{V}}}_N+\frac{1}{N}{\sum\limits_{j=i+1}^N}{\sum\limits_{i=1}^{N-1}\vec{v}_{ij}{\dot{\vec{v}}_{ij}}}$,
where
${\dot{\vec{V}}_N}=\frac{1}{N}{\sum\limits_{i=1}^N}\dot{\vec{v}_{i}}$;
$\vec{v}_{ij}=\vec{v}_i-\vec{v}_j$ are relative velocities.

By grouping the terms which determine bound and motion energies of
subsystems using relative velocities of elements, distances
between elements and velocities of centers of mass of subsystems
as variables we will obtain following equations with the help of
eqs. (d, e) [19]:

\begin{equation}
{Lm\vec{V}_L\dot{\vec{V}}_L+{\sum\limits_{j=i+1}^L}\sum\limits_{i=1}^{L-1}\{\vec{v}_{ij}
[\frac{{m\dot{\vec{v}}}_{ij}}{L}+\vec{F}_{ij}]\}=
-\sum\limits_{{j_K}=1}^K}\sum\limits_{{i_L}=1}^{L}\vec{v}_{i_L}
\vec{F}_{{i}_{L}{j}_{K}} \label{eqn7}
\end{equation}

\begin{equation}
{Km\vec{V}_K\dot{\vec{V}}_K+{\sum\limits_{j=i+1}^K}\sum\limits_{i=1}^{K-1}\{\vec{v}_{ij}
[\frac{{m\dot{\vec{v}}}_{ij}}{K}+\vec{F}_{ij}]\}=
\sum\limits_{{j_K}=1}^K}\sum\limits_{{i_L}=1}^{L}\vec{v}_{j_K}
\vec{F}_{{i}_{L}{j}_{K}} \label{eqn8}
\end{equation}

Left sides in the eqs. (7, 8) determine changes of subsystems'
energies $T_N^{tr}$ and $E_N^{ins}$, as a result of interaction of
subsystems. The first terms represent the change of kinetic energy
of subsystems' motion as a whole. The second terms describe
transformation of the bound energy. The right sides of the eqs.
(7, 8) describe the subsystems' interaction and determine the rate
of an energy exchange between subsystems.

Velocities of particles' of any subsystem can be presented as the
sum of velocities of the center of mass of the subsystem and their
velocities in relation to the center of mass. I.e.,
$v_i=\tilde{v_i}+V$. Then after grouping both parts of eqs. (7, 8)
in appropriate way, we will rewrite eqs. (7, 8) as:
\begin{equation}
L\vec{V}_L[m\dot{\vec{V}}_L+\vec{\Psi}]+{\sum\limits_{j=i+1}^L}\sum\limits_{i=1}^{L-1}\vec{v}_{ij}
[\frac{m\dot{\vec{v}}_{ij}}{L}+\vec{F}_{ij}]=\vec{\Phi}_L
\label{eqn9}
\end{equation}

\begin{equation}
K\vec{V}_K[m\dot{\vec{V}}_K-\vec{\Psi}]+{\sum\limits_{j=i+1}^J}\sum\limits_{i=1}^{J-1}\vec{v}_{ij}
[\frac{m\dot{\vec{v}}_{ij}}{K}+\vec{F}_{ij}]=\vec{\Phi}_K
\label{eqn10}
\end{equation}
Here $\vec{\Psi}=
{-\sum\limits_{{j_K}=1}^J}\sum\limits_{{i_L}=1}^{L}
\vec{F}_{{i}_{L}{j}_{K}}$; $\vec{v}_i=\tilde{\vec{v}_i}+\vec{V}$;
$\vec{\Phi}_L=
{-\sum\limits_{{j_K}=1}^J}\sum\limits_{{i_L}=1}^{L}\tilde{\vec{v}}_{i_L}
\vec{F}_{{i}_{L}{j}_{K}}$; $\vec{\Phi}_K=
{\sum\limits_{{j_K}=1}^J}\sum\limits_{{i_L}=1}^{L}{\tilde{\vec{v}}_{j_K}
\vec{F}_{{i}_{L}{j}_{K}}}$.

The eqs. (9, 10) determines the energy exchange between $L$ and
$K$ subsystems. We will call this equation as the {\it Equation of
Systems Interaction}.

As it follows from the right side of eqs. (9, 10), the change of
energy of $L$-subsystem as a result of its interaction with
$K$-subsystem is determined by velocities of motion of
$L$-subsystem's particles in relation to its center of mass and
potential interaction with particles of $K$-subsystem and vice
versa.

The first terms of the left side of eqs. (9, 10) determines the
change of the motion energy one subsystem as a wall in a potential
field of another subsystem. The second term determines subsystems'
bound energy change as a result of motion of particles one
subsystem in a field of particles of another subsystem.

Non-potentiality of forces of subsystems' interaction is caused by
transformation of kinetic energy of their motion not only into a
potential energy of subsystems as a whole, but also into bound
energy. Increase of a bound energy occurs as a result of decrease
of ordered motion of particles or decrease of velocity of the
relative motion of subsystems.

When $\vec{V}_L=0, \vec{V}_K=0$, energy of relative subsystems'
motion is absent. In this case the full systems' energy is equal
to the sum of bound energies of subsystems.

Let us explain, why bound energy is increasing only. Firstly we
take an equilibrium system. If we divide this system into two
subsystems by any way, they also will be equilibrium. Therefore
generalized forces and relative subsystems velocities are equal to
zero. It is clearly that in this case the spontaneous deviation
from equilibrium is possibly only in the case of spontaneous
occurrence of collective motion of particles which will lead to
occurrence of relative motion of subsystems. It is equivalent to
occurrence of momentum of a subsystem, which can arise only due to
the internal energy. But the law of conservation of momentum
forbids it.

In the second case we consider systems which consists of two
subsystems in motion to each other. Their relative velocity cannot
be increased due to their interaction energy as this energy is
determined by velocity relative motion itself. As well as in the
previous case the relative velocity of these subsystems cannot be
increased due to their bound energy.

Thus, the decreasing of relative velocities of subsystems is
possibly only. This decrease is caused by the fact that the work
of generalized forces increases internal energy of subsystems due
to kinetic energy of relative motion. The increasing of bound
energy is non-potential as a result of decrease of orderliness of
motion of particles, i.e. as a result of chaotization of vectors
of particles' velocities. As the inverse process is impossible,
generalized forces can decrease only.

In the case when velocities of particles' motion inside subsystems
can be neglected (hard body systems approach) the right side term
and the second term in the left side of Equation of Systems
Interaction are equal to zero. It is easy to see that Equation of
Systems Interaction in this case will be transformed to the
Newton's equations for two hard bodies:
$m\dot{\vec{V}}_L=-\vec{\Psi}$ and $m\dot{\vec{V}}_K=\vec{\Psi}$.

It is not so difficult to check, that these equations can be
obtained directly by summation of the Newton equations (c) for
particles of each subsystem. As the sum of the forces inside
subsystems will be zero, the generalized forces will become the
central subsystem-subsystem interaction forces. These forces will
be potential as changes of bound energy of subsystems does not
exist at such summation of the Newton equations (the total work of
forces of interaction of elements in the closed system will be
zero). Thus the Newton equations is follows from the Equation of
Systems Interaction when we neglect particles' motion in relation
of the center of masses of corresponding subsystems.

\section{Difference between particles dynamics and subsystems dynamics}

The Newton equation (c) can be treated as equation for particles'
interaction forces. The work of these forces determine
transformation of particles' kinetic energy to their potential
energy. This energy transformation takes place during transition
of system from one point of configuration space into another
[12,14]. Forces are determined by gradient of potential energy of
particles. Thus, forces and potential energy of particles are
completely determined by coordinates, and work of potential forces
along closed contour is equal to zero. It corresponds to
reversible dynamics.

And now we will consider Equation of Systems Interaction. From it
follows that in nonequilibrium systems kinetic energy of relative
subsystems motion is existed. This energy is connected with the
rate of regularity of particles motion of subsystems. The
regularity is determined by deviation from equilibrium of
subsystem velocities' distribution functions. As against Newton's
forces, the work of generalized forces will transform kinetic
energy of subsystem motion not only to the subsystem potential
energy as a whole, but also to bound energy. Because of such
transformation, the work of the generalized forces along the
closed contour in configuration space is not equal to zero.

The transition of subsystems' bound energy into kinetic energy of
subsystem is impossible. Really, this transition would be possible
only under condition of spontaneous occurrence of generalized
forces inside an equilibrium subsystem. But such occurrence means
infringement of spherical symmetry of distribution function of
elements velocities of equilibrium subsystems' in relation to the
center of mass. And it contradicts to the law of momentum
preservation.

Thus, Equation of Systems Interaction as against the Newton
equation describes process of transformation of systems' energy
which is caused not only by transformation of potential energy
into kinetic energy, but also by change of distribution function
of velocities of particles due to increase the rate of chaotic
motion of particles.

There is a question why the Newton equation is suitable for
description of particles' dynamics, but nevertheless it does not
determine systems' equilibration? Let us offer the following
answer. Dynamics of selected particles is unequivocally determined
by the equation of Newton. A motion of any particle is reversible.
But dynamics of subsystem is described by collective parameters
such as bound energy, generalized forces etc. These parameters
ambiguously depend on parameters of particles' motion. Such
ambiguity leads to occurrence of new collective systems' laws
which do not exist for separate particles. Let's consider for
example, velocity of motion of systems' centre of masses, which is
constant in the homogeneous space. This velocity is a collective
parameter of system. It is determined by the sum of velocities of
all particles of system. Therefore biunique conformity between
velocity of center of mass and particles' velocities does not
exist. The impossibility of increase of motion energy of an
equilibrium subsystem due to its bound energy is collective law
which determines its dynamics. Therefore, inspite of reversibility
of dynamics of separate particle, the subsystems dynamics can be
irreversible. Thus, irreversibility is a new property of systems
which is absent in dynamics of their particles. Occurrence of this
property within the frame of laws of classical mechanics becomes
possible owing to ambiguous dependence of dynamics' parameters of
subsystems on parameters which determine dynamics of individual
particles.

By exclusion of particles' potential interaction from Equation of
Systems Interaction with the help of eq. (1), we will obtain the
Equation of Systems Interaction for elastic disks  systems. It
means that the nature of irreversibility is identical both for
systems of elastic disks and for systems of potentially
interacting elements.

\section{Classical mechanics and thermodynamics}

It is possible to come to thermodynamics with the help of the eqs.
(7, 8). Actually, the right sides of these equations determine an
exchange of energy between subsystems due to their interaction.
The first term of the left side of each equation determines the
change of the motion energy of subsystem as a whole. In
thermodynamics this corresponds to mechanical work which is
carried out by external forces acting on subsystem from inside.
The second term of the left side corresponds to increase of bound
energy of a subsystem due to energy of relative motion of
subsystems. In thermodynamics this term corresponds to the change
of thermal energy of system.

It is easy to see the connection between eq. (7) and the basic
equation of thermodynamics [2, 3]: ${dE=dQ-{PdY}}$. Here, $E$ is
energy of a subsystem as a whole; $Q$ is thermal energy; $P$ is
pressure; $Y$ is volume according to common terminology,.

Change of energy of selected subsystem is due to the work carried
out by external forces. Therefore, the change of full energy of a
subsystem corresponds to $dE$.

The change of kinetic energy of motion of subsystem as a whole,
$dT^{tr}$, corresponds to the ${PdY}$ term. Really,
${dT^{tr}=\vec{V}d\vec{V}=\vec{V}\dot{\vec{V}}dt=\dot{\vec{V}}d\vec{r}=PdY}$
[2].

Let's determine, what term in Equation of Systems Interaction
corresponds to the change of bound energy. According to virial
theorem [14, 21], if potential energy is a homogeneous function of
second order of radiuses-vectors then
${\bar{E}^{ins}=2\bar{T}^{ins}=2\bar{U}^{ins}}$. Lines denote
averaging time. We have obtained above that bound energy,
${E^{ins}}$, increases due to contribution of ${T^{tr}}$ energy.
But the opposite process is impossible. Therefore change of $Q$
term corresponds to the change of ${E^{ins}}$, bound energy.

Let's consider the system to be near to equilibrium. If the
subsystem consists of ${N_m}$ elements then average energy of each
element can be written down as
${\bar{E}^{ins}={E}^{ins}/N_m=\kappa{T}_0^{ins}}$.

Now let us the bound energy be increased by ${dQ}$. According to
the virial theorem and keeping terms of the first order, we will
have: ${dQ\approx{T}_0^{ins}[d{E}^{ins}/{T}_0^{ins}]
={T}_0^{ins}[{d\vec{v}}/{\vec{v}_0}]}$, where ${\vec{v}_0}$ is the
average velocity of an element, and ${d\vec{v}}$ is its change.
For subsystems in equilibrium, we will have
${d\vec{v}/\vec{v}_0\sim{{d\Gamma_m}/{\Gamma_m}}}$, where
${\Gamma_m}$ is the phase volume of a subsystem, ${d\Gamma_m}$
will be increased due to increase of the subsystem energy at
${dQ}$ value. By keeping terms of the first order we will obtain:
${dQ\approx{T}_0^{ins}d\Gamma_m/\Gamma_m={{T}_0^{ins}}d\ln{\Gamma_m}}$.
According definition ${d\ln{\Gamma_m}=dS^{ins}}$, where
${S^{ins}}$ is subsystem entropy. So
${dQ\approx{T}_0^{ins}dS^{ins}}$ near equilibrium.

Let's consider the connection between generalized forces and
entropy. According to eq. (2) the entropy production in
non-equilibrium system is determined by transformation of kinetic
energy of subsystems motion into bound energy. Eventually relative
velocities of subsystems and generalized forces will comes to
zero. As a result the energy of relative motion of subsystems
completely transforms into bound energy and the system becomes
equilibrate. It means that energy of motion of subsystem was spent
on increase of entropy. Therefore the deviation of entropy from
equilibrium is determined by the following formula [19]:
\begin{equation}
{{\Delta{S}}={\sum\limits_{l=1}^R{\{{m_l}
\sum\limits_{k=1}^{m_l}\int{\sum\limits_s{{\frac{{F_{ks}}^{m_l}v_k}{E^{m_l}}}}{dt}}\}}}}\label{eqn11}
\end{equation}

Here ${E^{m_l}}$ is a kinetic energy of subsystem; ${m_l}$ is a
number elements in ${"l"}$ subsystem; ${R}$ is a number of
subsystems; ${s}$ is a number of external disks which were
collided with internal disk ${k}$; ${F_{ks}^{m_l}}$ is a force,
acting on $k$-disks; $v_k$ is a velocity of $k$- disk.

The integral in eq. (11) determines the work of ${F_{ks}^{m_l}}$
force, during the system's relaxation to equilibrium. In
equilibrium the energy of relative motion of subsystems and
generalized forces are equal to zero. I.e. the integral in eq.
(11) is determined by the energy of relative motion of subsystems.
It corresponds to phenomenological Clauses formula for entropy [2,
3, 18]. So eq. (11) will be in accordance with the eq. (5). Really
if ${E_l^{ins}\gg{T_l^{tr}}}$, than we have:
${dS=\sum\limits_{l=1}^R\frac{\partial{S_l}}{\partial{T_l^{tr}}}{dT_l^{tr}}}$.
It corresponds to eq. (11). Both in eq. (5) and eq. (11) entropy
increasing is determined by change of the energy of relative
motion of subsystems.

Thus, eq. (11) connects dynamic parameter which is generalized
force acting on a subsystem, with entropy which is a thermodynamic
parameter. I.e. eq.(11) establishes connection between parameters
of classical mechanics and thermodynamic parameters. The deviation
of system from equilibrium is characterized by the ratio between
energy of relative motion of subsystems and full energy of system.

\section{Conclusion}

As a result of investigation of a hard-disks system, the necessity
of expansion of the canonical formalism of Hamilton for
description of irreversible dynamics has been found out. This
expansion consists in adaptation of the canonical formalism of
Hamilton to non-potentially interacting systems. As a result of it
expansion the generalized Liouville equation was obtained. Based
on the this equation it was obtained that non-potentiality of
generalized forces is the necessary condition for irreversible
dynamics.

Using a method of splitting of nonequilibrium systems into the
equilibrium subsystems and basing on generalized Liouville
equation, the problem of irreversibility was transformed to the
problem concerning character of behavior of generalized forces. In
order to solve this problem the equation of subsystems interaction
was obtained. The non-potentiality of the forces acting between
subsystems and their tendency to zero has been found from this
equation.

Following explanation of irreversibility can be offered. In spite
of potentiality of fundamental forces, generalized forces are
non-potential. The work of these forces  irreversibly  transforms
energy of relative motion of subsystems to their bound energy. It
is so because in accordance with the law of conservation of
momentum, velocity of relative motion of a subsystem cannot be
increased due to bound energy. Therefore these forces and relative
velocities of subsystems are being decreased. The process of
decreasing of the relative velocities of subsystems and increasing
of bound energy is performed due to increase of chaotization of
vectors of particles' velocities. Thus irreversible decrease of
energy of relative motion of subsystems due to its transformation
to bound energy determines is essence of the irreversibility
mechanism and of the equilibration process.

Since the non-potentiality of generalized forces is the essence of
irreversibility, it is not possible to explain irreversibility
within the framework of canonical Hamilton equations because of
their inapplicability for description of dynamics of systems with
non-potential forces [12].

The first law of thermodynamics follows from the subsystems'
interaction equation. This equation determines the transformation
of the work of subsystem-subsystem interaction forces to the bound
energy and the energy of subsystem motion.

The content of the second law of thermodynamics is determined by
irreversible transformation of energy of motion of subsystems to
their bound energy since amount of energy of subsystems' motion,
which is transformed to bound energy corresponds to increase of
entropy.

The offered mechanism of irreversibility determines the horizon of
probabilistic description of systems' dynamics. In accordance with
this mechanism it is impossible to use probabilistic description
for nonequilibrium systems if characteristic time of equilibration
due to generalized forces is less than probabilistic time of the
occurrence of the system's nonequilibrium state.

Thus, the results of irreversibility investigation given here
confirm an opportunity of a substantiation of thermodynamics
within the frame of classical mechanics.

\smallskip

\end{document}